\documentclass[a4paper,11pt]{article}
\pdfoutput=1 

\usepackage{jheppub} 
\usepackage[T1]{fontenc} 

\usepackage{amssymb}

\usepackage[utf8]{inputenc} 
\usepackage{graphicx}

\usepackage{braket}
\usepackage[vcentermath]{youngtab}

 \usepackage{fancyhdr}

\usepackage{mathrsfs}
\usepackage[T1]{fontenc}
\usepackage{setspace}
\usepackage{amsfonts}
\usepackage{amssymb}
\usepackage{amsmath}
\usepackage{epsfig}
\usepackage{latexsym}
\usepackage{color}
\usepackage{nicefrac}
 \usepackage{slashed}
 \usepackage{multirow}
 \usepackage{comment}
 \usepackage{soul}
 \usepackage{hyperref}
\usepackage{slashed}
\usepackage{simpler-wick}
\usepackage{array}
\usepackage{tabu}
\usepackage{tcolorbox}
\usepackage{subcaption}
\usepackage{cancel}
\usepackage[dvipsnames]{xcolor}

\usepackage{tikz}
\usetikzlibrary{arrows.meta}

\newcommand{\ea}[1]{
\begin{align}
#1
\end{align}
}

\newcommand{\beq}{\begin{eqnarray}}
\newcommand{\eeq}{\end{eqnarray}}

\newcommand{\p}{{\cal P}\exp}

\newcommand{\bmp}{\noindent\begin{minipage}{16cm}}
\newcommand{\emp}{\end{minipage}\vskip 7mm} 

\newcommand   \cO {\mathcal{O}}

\renewcommand \l  {\lambda}
\newcommand   \<   {\langle}
\renewcommand \>   {\rangle}

\newcommand*{\del}{\mathop{\mathrm{{}\partial}}\mathopen{}}

\newcommand{\ee}{\end{equation}}
\newcommand{\be}{\begin{equation}}
\newcommand{\bea}{\begin{align}}
\newcommand{\eea}{\end{align}}
\newcommand{\eps}  {\epsilon}

\renewcommand{\d}{\textrm{d}}

\def\lsim{\mathrel{\rlap{\lower4pt\hbox{\hskip1pt$\sim$}}
    \raise1pt\hbox{$<$}}}                
\def\gsim{\mathrel{\rlap{\lower4pt\hbox{\hskip1pt$\sim$}}
    \raise1pt\hbox{$>$}}}                
    
    \def\be{\begin{equation}}
\def\ee{\end{equation}}
\def\ba{\begin{eqnarray}}
\def\ea{\end{eqnarray}}

\def\del{\partial}

\def\r{\rho}
\def\a{\alpha}

\def\b{\beta}

\def\d{\delta}
\def\D{\Delta}
\def\e{\epsilon}

\def\p{\pi}

\def\c{\chi}

\def\l{\lambda}

\def\j{\psi}

\def\co{\mathcal O}
\def\del{{\partial}}

\def\cf{{\cal F}}
\def\ch{{\cal H}}

\def\cb{{\cal B}}
\def\cc{{\cal C}}

\def\IR{\relax{\rm I\kern-.18em R}}

\def\IR{\relax{\rm I\kern-.18em R}}
\def\IL{\relax{\rm I\kern-.18em L}}

\def\inv{^{\raise.15ex\hbox{${\scriptscriptstyle -}$}\kern-.05em 1}}

\def\cO{{\cal O}}

\title{\boldmath Quantum chaos from time-dependent CFT states  at large central charge }

\author{Shaun D. Hampton, Pantelis Panopoulos, Gideon Vos}

\affiliation{Korea Institute for Advanced Study, 85 Hoegi-ro, Dongdaemun-Gu, Seoul 02455, Korea}

\emailAdd{sdh2023,ppanopoulos,gideonvos@kias.re.kr}

\abstract
{We provide a first study of the behavior of four probes on heavy, time-dependent background states of conformal field theories  at large central charge. Employing the monodromy method and including subleading corrections to the identity block, we compute the correlation function  of two  light and two middle-heavy operators on  such a heavy background, by switching on a small interaction between the light and middle-heavy sectors. Summing   the discontinuity contributions  and analytically  continuing from Euclidean to Lorentzian signature, we show that the out-of-time-ordered correlator (OTOC) exhibits maximal chaos,  that is, the Lyapunov exponent saturates the MSS bound.}

\preprint{KIAS-P26087}

\begin{document}

\maketitle
\flushbottom

\section{Introduction}
\label{sec:introduction}

Within many-body classical systems chaotic behavior  is a commonplace and well-defined phenomenon.  
In quantum systems however, we instead  have only a number of well-studied diagnostics for chaos, which include the out-of-time-ordered correlator (OTOC), spectral form factor, and level repulsion in matrix mechanics.   The  OTOC is defined as a four-point function and it detects chaos through the Lyapunov exponent. This has  first been studied    in  two dimensional  conformal field theories at large central charge in \cite{Roberts:2014ifa,Maldacena:2015waa,Jackson:2014nla}. These calculations were initially inspired by holography where the chaotic behavior was considered in similar set ups in \cite{Shenker:2013pqa,Roberts:2014isa} which  in turn was inspired by  \cite{Sekino:2008he}. An effective hydrodynamic approach of chaos in field theory  and holography  was undertaken in \cite{Blake:2017ris,Blake:2018leo} while an effective action based on reparametrization modes was considered in  \cite{Haehl:2018izb,Nguyen:2021jja}.

In a  quantum chaotic system we recognize two important time-scales,  the dissipation time $t_d$ and the scrambling time $t_*$. The dissipation time is the timescale where the thermal  two-point function of an operator $\co(t)$ decays exponentially in the following way,
\be
\<\co(t)\co(0)\>_\b\sim e^{-t/t_d}\,,
\ee
where $\<\dots\>_\b$ a thermal correlator, and  physically it measures the relaxation time under  small perturbations. Therefore the two-point function does not give an actual diagnostic for chaos. The OTOC of two operators $\co_M$ and $\co_L$  defined by 
 \be
 \<\co_M(t)\co_L(0)\co_M(t)\co_L(0)\>_\b\,,
 \ee
  is contained within  the expectation value of the commutator squared
\begin{equation}\label{C}
   C(t) =  -\langle[\co_M(t),\co_L(0)]^2\rangle_\b.
\end{equation}
 In our framework, we assign to  $\co_M(t)$ and $\co_L(0)$  different  light weights and call them  {\it middle heavy} and {\it light} operators respectively.  The system exhibits  chaos whenever \eqref{C} behaves as,
\begin{equation}
     C(t) \sim \langle \co_L\co_L\rangle_\beta \langle \co_M\co_M\rangle_{\beta} \left(1- \frac{u}{c}e^{\lambda t}\right),
\end{equation}
where  $\l$ is the quantum Lyapunov exponent and $u$ is an unspecified constant of order $\mathcal{O}(c^0)$. Under general principles it was shown in \cite{Maldacena:2015waa}  that the Lyapunov exponent is bounded,  
\be
\l\leq\frac{2\p}{\b}\,,
\ee
and the  value $\l=\frac{2\p}{\b}$ is the MSS bound  saturated by  maximally chaotic systems. Physical implications of this bound were addressed in \cite{Hartman:2015lfa,Chandorkar:2021viw} and $1/c$ corrections  in \cite{Fitzpatrick:2016thx}.

In this work we consider time-dependent pure states. The novelty of studying such systems is related to the fact that, in principle, they correspond to non-equilibrium systems and consequently not to thermal states. However it was shown in \cite{Fitzpatrick:2015zha,Banerjee:2018tut,Vos:2025voq} that there exists a long-lived equilibrium mode which drives effective thermalization. Specifically, in the time-dependent case \cite{Banerjee:2018tut,Vos:2025voq},  at time-scales larger than the scale set by the physical system size the two-point expectation value of light operators $\co_L$ in the leading order behaves as
\begin{equation}
    \langle \mathcal{H}| \co_L(0)\co_L(t)|\mathcal H\rangle \sim e^{-2\pi h_L t/\beta},
\end{equation}
i.e. it takes the form of a Lorentzian thermal two-point function with an effective inverse temperature $\beta$ given by
\begin{equation}
    \beta = \frac{\pi}{|\alpha|}\,,
\label{temperature}
\end{equation}
where $\alpha$ is the Floquet exponent. 
The goal of this paper is to continue to four-point expectation values. As  \eqref{C} indicates, the natural object that captures chaos for a time-dependent state should be
\begin{equation}
   {C}_\ch(t) =  -\langle \mathcal H|[\co_M(t),\co_L(0)]^2|\mathcal H\rangle\,.
\end{equation}
The extra weight causes a small indirect coupling between $\cO_M$ and $\cO_L$.  As it will be shown, the alternating terms in the commutator squared acquires the general form
\begin{equation}
     C_\ch(t) \propto \,\langle \mathcal H|\co_L\co_L|\mathcal H\rangle \; \langle \mathcal H|\co_M\co_M|\mathcal H\rangle \left(1 - \frac{h_L h_M}{c} \cb_-^{(1)}(t)e^{2|\alpha|t}\right)\,,
\label{ourOTOC}
\end{equation}
where $\cb_-^{(1)}(t)$ is a bounded function of time.
Comparing with \eqref{C} we show that the Lyapunov exponent saturates the MSS bound, namely 
\be
\l=2|\a|\,.
\ee

The rest of the paper is organized as follows: In section \ref{sec:LorentzianCFT} we define the three classes of time-dependent operators that we are working with and separate between  background and probes. We also give a detailed account of the structure of Lorentzian correlation functions and the analytic continuation from Euclidean to Lorentzian space. In section \ref{Conformal-blocks}, after reviewing the  essentials of the monodromy method and its extension to subleading corrections from heavy-light perturbation theory, we reproduce the two-point function of light operators. Then, we evaluate the correlation function of four light operators in a heavy background at large $c$. In section \ref{sec:quantumchaos}, after defining the four-point function consisting of two middle-heavy and two light operators we extract the Lyapunov exponent and show that it saturates the MSS bound. Finally, in section \ref{sec:discussion} we summarize  our results and discuss  some future directions.

\section{Two-dimensional   CFTs on the Lorentzian cylinder }

\label{sec:LorentzianCFT}

In this section we are focused in the general structure  of Lorentzian and Euclidean correlators of time-dependent states. We start by introducing   three different classes of time-dependent operators that play the role of background heavy states and probes in  the CFT. Then, after highlighting some  of the essential properties  of the Lorentzian correlators and their relation to Euclidean ones, we set our coordinate conventions and the two level time ordering.

\subsection{Time-dependent states}

At CFTs with $c\gg1$ we can separate  operators $\co_\D$ to according the value of their scaling dimensions $h_\D$ defined by $[L_0,\co_\D]=h_\D\co_\D$\,. For the purpose of our work  we distinguish between heavy operators $\co_H$,  middle-heavy operators $\co_M$ and light operators $\co_L$, with scaling dimensions denoted by $H$, $h_M$ and $h$ respectively, characterized by the following conditions,
\begin{tcolorbox}[colback=yellow!7, colframe=blue!60!black]
\be\label{operator-classes}
   \co_H:\,\,  \lim_{c\rightarrow \infty} \frac{H}{c} = \text{finite}, \qquad \co_M:\,\,\lim_{c\rightarrow \infty} \frac{h_M}{c} = \zeta,\qquad \co_L:\,\lim_{c\rightarrow \infty} \frac{h}{c} = 0,
\ee
\end{tcolorbox}
where $\zeta$ is as small intermediate value satisfying
\begin{equation}
  \frac{1}{c}\ll   \zeta \ll 1.
\end{equation}
Throughout this work we consider heavy background states $|\mathcal{H}\rangle$,  defined as a product of $\mathcal{O}(1)$ number $n$ of heavy operators  acting on the CFT vacuum $|0\>$, under the condition $n\ll c$, namely\footnote{The set of states $\{|\mathcal{H}\rangle\}$ should not be thought of as a linear space, i.e. $|\mathcal{H}_1\rangle + |\mathcal{H}_2\rangle \notin \{|\mathcal{H}\rangle\}$.}
\begin{equation}\label{Heavy-states}
    \left\{|\mathcal{H}\rangle\right\} = \left\{ \co_{H_1}(x_1)...\co_{H_n}(x_n)|0\rangle\right\},\quad\qquad n\ll c.
\end{equation}
In principle,  heavy operators are complicated many-body objects in the degrees of freedom of the CFT and in our setting we  can think of them as the two dimensional  analog of single-trace operators in Yang-Mills theory. The fact though, that we have not imposed any conditions on the states \eqref{Heavy-states}, indicates the genericity for the mechanism we will discuss.  However, the special feature of these states is that they are not eigenstates of the Hamiltonian and therefore are genuinely {\it dynamical time-dependent} states. This property can be deduced by considering  the normalized quantum uncertainty of the $L_0$ on $|\mathcal{H}\>$. For a general state $|\Psi\>$ it is defined as
\be\label{uncertainty}
\frac{(\D L_0)^2}{\<\Psi|\Psi\>}=\frac{\langle \Psi|L_0^2|\Psi\rangle}{\<\Psi|\Psi\>} - \frac{\langle \Psi|L_0|\Psi\rangle^2}{\<\Psi|\Psi\>}\,.
\ee
Evaluating \eqref{uncertainty} on  a Hamiltonian eigenstate, it necessary vanishes. Let us  now consider $|\Psi\>=|\mathcal H\>$.  Using the Ward identities \eqref{conformal-WI}, the energy expectation value is given by 
\begin{equation}
        \langle \mathcal H|L_0|\mathcal H\rangle = \left(\sum_i^n H_i + x_i\partial_{x_i}\right) \langle \mathcal H|\mathcal H\rangle = \mathcal O(c).
\end{equation}
Due to some cancellations among the quadratic terms, the energy uncertainty \eqref{uncertainty} becomes 
\begin{equation}\label{H-uncertainty}
    \frac{\langle \mathcal H|L_0^2|\mathcal H\rangle - \langle \mathcal H|L_0|\mathcal H\rangle^2}{\langle \mathcal H|\mathcal H\rangle} = -\left(\sum_{ij} 2x_i H_j +n \sum_i x_i\right)= \mathcal O(c).
\end{equation}
Therefore, the non-zero nature of \eqref{H-uncertainty} in the energy decomposition confirms that these are not energy eigenstates. Moreover, an important observation  is that as $c$ becomes larger, the expectation value of $L_0$ grows linearly with $\mathcal O(c)$, whereas the standard deviation only grows as $\mathcal O(\sqrt{c})$. From this we conclude that in relative terms the spectral decomposition of the states in the $L_0$-eigenbasis becomes effectively more sharply peaked as $c$ increases and   the standard deviation never becomes negligible. This  extra evidence  indicates that the states we are considering are time-dependent pure states\footnote{It seems natural to consider them holographically as states corresponding to heavy multi-body initial states. For more details see \cite{Hulik:2016ifr,Anous:2016kss} and  the discussion in section \ref{sec:discussion}.}.

\subsection{Analytic structure of Lorentzian correlators}

Having defined the states we are considering on the background, we now discuss the analytical structure of the Lorentzian correlators which is not affected by the choice of state. This structure  is fairly intricate since in principle  Lorentzian correlators are multivalued functions of their operator locations with branch points running along the lightcones of the operator insertions. Generally, each Riemann sheet of this multivalued function corresponds to a different operator ordering \cite{Hartman:2015lfa, Simmon-DuffinTasi, Kundu:2025jsm}. In more formal settings each of these operator orderings corresponds to a separate Wightman function of the Lorentzian theory. For generic quantum field theories the reconstruction theorem \cite{Osterwalder:1973dx,Osterwalder:1974tc} states that all of the Wightman functions corresponding to different orderings, can be obtained as analytic continuations from a Euclidean correlator  \cite{Simmon-DuffinTasi}.

 The aforementioned branch points come with branch cuts that run in the imaginary time direction. The physical interpretation of these branch cuts is that every time a lightcone is encountered a choice has to be made: the lightcone is circumvented by moving briefly into the complex plane and  at this stage we have to decide  whether the point is circumnavigated with or without crossing the branch cut.  This choice corresponds to whether the operator ordering is exchanged or not.

In 2d CFTs the analytic continuation is particularly clear. Our operator's locations are defined by two coordinates $\mathcal O(z,\bar{z})$ and we view these coordinate labels $z,\bar{z}$ as independent complex variables. In this case both Lorentzian correlators and Euclidean correlators are obtained by considering two different subsurfaces of the two-complex dimensional plane \cite{Kundu:2025jsm}. We can either take the condition $\bar{z}=z^*$ and obtain Euclidean correlators or take $z,\bar{z}\in \mathbb{R}$ to hold for all the operators inside a correlator instead and find a Lorentzian correlator. Restricted to the Euclidean section correlation functions are single-valued functions of $z$.  Note that this is not  true for the conformal blocks that make up the correlators.

There is an elegant argument \cite{Hartman:2015lfa} from which we can now see that the different Wightman functions are analytic continuations of each other. If the Wightman functions are analytic functions of time, then their times can be dialed to zero at which point the Lorentzian and Euclidean sections coincide. Hence, every Wightman function of a correlator shares the same Euclidean correlator as a boundary function of their time arguments. Euclidean correlators are analytic on their domain outside of operator coincidences and thus direct application of the edge-of-the-wedge theorem shows that there exists an analytic extension from one Wightman function into the next. An overview of the correlator's behavior  is displayed in table \ref{multivaluednesstable}.

Note that while the correlator when restricted to the Euclidean section is single-valued, it does have branch cuts. Whenever $z$ crosses a branch cut and picks up a branch discontinuity there is a simultaneous branch cut crossing by $\bar{z}$ with a branch discontinuity which cancels that of the holomorphic sector. That this has to occur can be seen from the conformal block decomposition, where term-by-term the contributions factorize into products of holomorphic and anti-holomorphic conformal blocks \cite{Kundu:2025jsm}, namely
\begin{equation}
    \mathcal{G}(z,\bar{z}) = \sum_{h,\bar{h}} C_{h,\bar{h}} \mathcal{F}_h(z) \bar{\mathcal{F}}_{\bar{h}}(\bar{z}).
\end{equation}
\begin{table}[]
    \centering
    \begin{tabular}{c|c|c}
         & Branch points & Structure \\
         \hline 
       Unconstrained correlator  & \checkmark & multi-valued \\
       \hline
       Euclidean correlator $\bar{z}=z^*$ & \checkmark & {\it single-valued} \\
       \hline
       Lorentzian correlator $z,\bar{z}\in \mathbb{R}$ & \checkmark  & multi-valued \\
       \hline
       Conformal block & \checkmark & multi-valued \\
    \end{tabular}
    \caption{The only single-valued correlator is the Euclidean one. All Lorentzian correlators are always multivalued with branch-points running across the light-cone and result from the single Euclidean correlator under a particular choice of the analytic continuation.}
    \label{multivaluednesstable}
\end{table}

\subsection{Complexified coordinates and two levels of time ordering}

As specified in the introduction, the goal of this paper is to compute the OTOC on a heavy time-dependent background state $|\mathcal H\rangle$ . The starting point is the four point function
\begin{equation}
     C_{\mathcal{H}}(t) = -\langle \mathcal H|[\co(t),\co(0)]^2 |\mathcal H\rangle\,.
\end{equation}
defined on a time-dependent heavy background. Expanding ${C}(t)$, we obtain two types of correlators
\begin{equation}\label{correlators}
 \langle \mathcal H| \co(0) \co(t) \co(t) \co(0) |\mathcal H\rangle,\qquad   \langle \mathcal H| \co(t) \co(0) \co(t) \co(0) |\mathcal H\rangle\,,
\end{equation}
all measured against a reference time $t=0$ by taking into account the fact that the heavy state $|\mathcal H\rangle$ is itself time-dependent as well. Hence, we have two levels of time-ordering to consider. The first correlator in \eqref{correlators} will contribute trivially to $\mathcal{C}(t)$,  while $\<\ch|\co(0)\co(t)\co(0)\co(t)|\ch\>$ which is internally out-of-time order. Note also that  the generic expectation values due to the time dependence of the background, $\langle \mathcal H(t)|\co|\mathcal{H}(t)\rangle$ are out-of-time order as well. 

The strategy to evaluate the correlators in \eqref{correlators} in the Lorentzian signature will be the following. In all cases we analytically continue from the same correlator defined in Lorentzian space
\begin{equation}
    \langle \mathcal H| \co_M(t) \co_L(0)\co_L(0) \co_M(t)|\mathcal H\rangle,
\end{equation}
which  is  declared to be the principle sheet of the correlator. The reason for this will be explained in more detail in section \ref{sec:quantumchaos}.  Concisely,   it is useful to think of the Hermitian operators $\co_M(t)$ as producing a perturbed version of the state $|\mathcal H\rangle$, i.e.
\begin{equation}
    \langle \mathcal H| \co_M(t) \co_L(0)\co_L(0) \co_M(t)|\mathcal H\rangle = \langle \tilde{\mathcal H}|\co_L(0)\co_L(0)|\tilde{\mathcal{H}}\rangle,
\end{equation}
where
\begin{equation}\label{H-tilde}
    |\tilde{\mathcal H}\rangle = \co_M(t)|\mathcal H\rangle.
\end{equation}
This approach is conventional \cite{Simmon-DuffinTasi} and has been utilized in the past in similar context{\color{blue}{s}} \cite{Banerjee:2018tut,Vos:2025voq}. We will give to all the time-coordinates small imaginary parts $\eps_i$ and set the real Lorentzian times to zero. The result is a  correlator on the Euclidean section. The relative magnitudes of the imaginary times $\eps_i$ are chosen in such a way that the resulting Euclidean correlator is time-ordered in Euclidean time.

\begin{figure}
	\centering
		\includegraphics[scale=0.8]{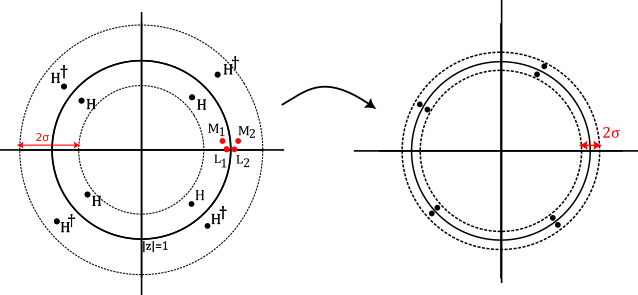}
	\caption{A diagram of the radial plane and the relative orderings of the heavy, middle-heavy and light operators. In order to ensure that the identity block dominates the Euclidean correlator, the annulus of width $2\sigma$ that contains all the operator insertions is taken to be small while maintaining the relative orderings of all the operators.}
	\label{operatororderingdiagram}
\end{figure}

As a conventinal frame, we express the Lorentzian operator positions in lightcone coordinates $x = t +\phi$ and $\bar{x}=t -\phi$, with time coordinate $t$ and {\it periodic spatial} coordinate $\phi$. Consider the operators $\co_i(x_i,\bar{x}_i)$ inside the correlators, where $x_i = t_i + \phi_i$ and $\bar{x}_i = t_i - \phi_i$. We will now complexify the time coordinates
\begin{equation}\label{t-complex}
    t_i \rightarrow t_i + i\eps_i,
\end{equation}
and subsequently send  $t_i$ to zero, $t_i\to0$. Additionally, we will express the correlator in terms of radial coordinates $z=e^{ix}$ and $\bar{z}=e^{i\bar{x}}$. Plugging  \eqref{t-complex} in $(z,\bar z)
$ components we get 
\begin{equation}
    z=e^{-\eps + it +i\phi}, \qquad \bar{z} = e^{-\eps + it -i\phi}.
\end{equation}
Notice that after taking $t\rightarrow 0$  we end up on the Euclidean section $\bar{z}=z^*$, with a Euclidean time-coordinate $\tau = -\eps$. Also since the Hermitian operators satisfy the conjugation rule $O(z,\bar{z})^{\dagger}=O(1/\bar{z},1/z)$, we parametrize the Euclidean correlator with the following $\eps$-prescription (we  suppress the spatial coordinate for brevity),
\begin{equation}
    \langle \cO_{H_1}(-\eps_1)....\cO_{H_n}(-\eps_n)\;\; \cO_M(-\eps)\cO_L(0)\cO_L(0)\cO_M(\eps)\;\; \cO_{H_n}(\eps_n)...\cO_{H_1}(\eps_1)\rangle,
\end{equation}
satisfying the inequalities
\begin{equation}
    \eps< \eps_n < ... <\eps_1.
\end{equation}
The positions of the operators are displayed schematically in Figure \ref{operatororderingdiagram}. The heavy operators are assumed to be restricted around a thin annulus centered around the unit circle \cite{Anous:2017tza} which ensures that the identity block is a reasonable approximation to the full Euclidean correlator.

\subsection*{Two levels of time ordering}
As mentioned above there are two distinct operator ordering subtleties that have to be dealt with. Firstly, if we want to compute the Lorentzian correlators 
\begin{equation}\label{orderings}
    \langle \mathcal{H}|\cO_L(0)\cO_M(t)\cO_L(0)\cO_M(t)|\mathcal H\rangle, \qquad \langle \mathcal{H}|\cO_M(t)\cO_L(0)\cO_M(t)\cO_L(0)|\mathcal H\rangle\,,
\end{equation}
we need a prescription to obtain these orderings from the Euclidean {\it principal} correlator,
\begin{equation}\label{principal-correlator}
    \langle \mathcal{H}|\cO_M(t)\cO_L(0)\cO_L(0)\cO_M(t)|\mathcal H\rangle,
\end{equation}
but additionally as mentioned, we need to encompass the inherent out-of-time ordering of expectation values of time-dependent states $\langle \mathcal H(t)| \cO ... \co |\mathcal H(t)\rangle$.

Let us first discuss how the two OTOC correlators are obtained from \eqref{principal-correlator}. As was mentioned earlier, ordering in Lorentzian spacetime is related to the lightcones of the operator insertions. Every point on the lightcone is a branchpoint from which a cut appears. Every time an operator is moved across the lightcone of another operator a choice has to be made. Recall that, we can either go around the branch point with or without crossing the branch cut. Crossing the branch cut exchanges the interpretation of the operator ordering within the Lorentzian correlator. 

Our starting point is the conformal block of the Euclidean correlator. From general principles we expect there to be branch points located at the operator locations with cuts running between two OPE contracted operators. This is confirmed by explicit computation. If we define the normalized light operator dependence of the Euclidean identity block as
\begin{equation}
    \mathcal{F}_{h=0}(z_1,z_2) = \frac{ \langle \mathcal H|\co_M\co_L(z_2)\co_L(z_1)\co_M|\mathcal H\rangle_{h=0}}{\langle \mathcal H|\co_M\co_M|\mathcal H\rangle_{h=0}}, 
\end{equation}
then, as will be discussed in the next section, to leading order in large $c$ we can express the normalized light operator dependence of the Euclidean identity block as
\begin{equation}
    \mathcal{F}_{h=0}(z_1,z_2) = \frac{1}{\left(\rho^M_+(z_1)\rho^M_-(z_2)-\rho^M_+(z_2)\rho^M_-(z_1)\right)^{2h_L}}.
\end{equation}
where  the functions $\rho^M_{\pm}(z)$ are two linearly independent solutions of the linear differential equation
\begin{equation}
    \frac{d^2}{dz^2}\rho^M_{\pm}(z) + \frac{6}{c} T_M(z) \rho^M_{\pm}(z) = 0.
\end{equation}
and $T_M(z)$ is the stress tensor expectation value of the heavy and middle-heavy sector.  
Quite importantly, the function $T_M(z)$ has second-order poles at the locations of both the heavy and middle-heavy operators\footnote{We elaborate further on this  fact in section \ref{sec:quantumchaos}.} and due to  the general theory of linear differential equations the solutions will have branch points at the operator locations\footnote{This can easily be seen for example by performing an asymptotic approximation of the solution around one of the double poles, the leading term of this expansion will be a power law with a generically non-integer power which is fixed by the coefficient of the double pole.}.

\begin{figure}[htbp]
    \centering
    \includegraphics[width=0.35\textwidth]{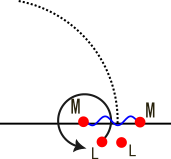}
    \hspace{10mm}
    \includegraphics[width=0.35\textwidth]{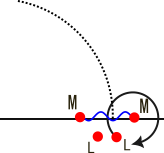}
    \caption{\textbf{\textit{Left:}} The exact path $\mathcal{C}_1$ the operator located at $z_1$ takes in order to obtain the $\cO_M \cO_L\cO_M\cO_L$-ordering. The point $z_1$ is taken a full counter-clockwise loop across the branch cut by looping around the middle-heavy operator located at $w_2$. \textbf{\textit{Right:}} The contour $\mathcal{C}_2$, the point $z_2$ takes a full clockwise loop across the branch cut by looping around the middle-heavy operator located at $w_2$.}
    \label{orderingfigure}
\end{figure}

In order to produce the term with the ordering $\cO_M\cO_L\cO_M\cO_L$ in \eqref{orderings}, we have to move the earliest operator in \eqref{principal-correlator} through a branch cut associated to the earliest $\co_M$ operator and subsequently return the operator to its original position. Since there is only one branch cut corresponding to that operator the only option is a closed loop around the operator. This is displayed in the left image in Figure \ref{orderingfigure}. Similarly, the term $\cO_L\cO_M\cO_L\cO_M$ in \eqref{orderings} is obtained by starting from \eqref{principal-correlator}, by moving the latest $\cO_L$ operator a full circle around the latest $\cO_M$ in operator and picking up the subsequent branch discontinuity as  shown in the right image of Figure \ref{orderingfigure}. 

The second step in the time-evolution process involves the analytic continuation of  the operators $\cO_M(t)$ to Lorentzian time. It is not immediately obvious which trajectory on the complex plane is the correct  one so that we  finally end up on the correct Riemann sheet. This problem was considered in \cite{Banerjee:2018tut} and a similar approach was described in \cite{Simmon-DuffinTasi}. Here we will follow closely  \cite{Vos:2025voq} which was inspired by \cite{Kundu:2025jsm}. This involves considering the trajectory on the Lorentzian cylinder.

\begin{figure}
	\centering
		\includegraphics[scale=1]{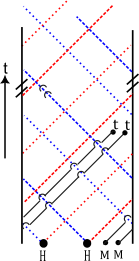}
	\caption{An example of a  lightcone diagram for an initial state consisting of two heavy operators. Their left- and right-moving lightcones are indicated respectively with blue and red dashed lines. The middle-heavy operators are evolved along the indicated trajectory. In the diagram a single `lightcrossing time' is indicated, i.e. the amount of parameter time it takes for the middle-heavy operators to return to their initial position.}
	\label{lightcrossingdiagram}
\end{figure}

\begin{figure}
	\centering
		\includegraphics[scale=1]{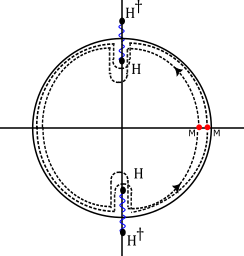}
	\caption{The diagram on the Euclidean radial plane of  the  trajectory in the Lorentzian spacetime diagram \ref{lightcrossingdiagram}. The wiggling lines between the pairs of heavy operators indicate the branch cuts between the heavy operator locations in the Euclidean conformal block. The parametrization of the middle-heavy operators is given in \eqref{radialplanetrajectory}. Note that due to the factor of $2$ in the exponent of \eqref{radialplanetrajectory} the `lightcrossing'  in Figure \ref{lightcrossingdiagram}, corresponds to two full rotations of the middle-heavy operators around the radial plane.}
	\label{MiddleHeavyTimeEvolution}
\end{figure}

Time-evolved operators are required to move along a trajectory on the Lorentzian cylinder depicted in Figure \ref{lightcrossingdiagram}. To avoid clutter we will consider a heavy state that  consists only of two operators and we only draw the lightcones of the heavy operators. We also parametrize this trajectory with a parameter $s$ and  for our operator to end up at time $t$, we assume that $s\,\in [0,t]$\footnote{To be more precise, we will assume that we can analytically continue the real parameter $s$ to a thin strip on the complex plane surrounding the real axis. The reason for this treatment is because the $\co_M$ operators in the Euclidean plane have to be separated with imaginary parts $\pm i\e$ which encode their relative ordering.}, namely
\begin{equation}\label{trajectory}
    t(s) = s, \qquad \phi(s) = \phi_0+s.
\end{equation}
At the Euclidean level we have only computed the holomorphic identity block and ignored the anti-holomorphic sector. For this reason it is practical to consider a trajectory on the Lorentzian cylinder that leaves the coordinate $\bar{z}$ fixed. This is legitimate since as it was argued in \cite{Kundu:2025jsm}, all paths lead to the same answer as long as  the endpoints remain the same. Implementing \eqref{trajectory} in radial coordinates, it leads to
\begin{equation}
    z(s) = e^{ix(s)} = e^{2is + i\phi_0}, \qquad \bar{z}(s) = \bar{z}.
\label{radialplanetrajectory}
\end{equation}
Subsequently, turning on Lorentzian time corresponds to performing a series of counterclockwise loops centered around the origin as shown in Figure \ref{MiddleHeavyTimeEvolution}. Note the importance of the factor 2 in the exponent. As $t$ is taken to be (a multiple of) $t=2\pi$ we loop around the complex plane twice and hence pick up the branch cut of each heavy operator pair twice. Moving on the Lorentzian cylinder, every single `lightcrossing time', i.e. the time it takes for a lightlike signal to propagate a full loop around the cylinder, we cross every heavy operator lightcone twice as well. This synchronicity suggests that this strategy provides the correct prescription.

\section{Correlation functions in Euclidean space}

\label{Conformal-blocks}

In the previous section we discussed the systematics of the analytic structure of Lorentzian correlators along with correct prescriptions when moving from Euclidean to Lorentzian space. This section is focused on the Euclidean dynamics. After reviewing the essentials of the  monodromy method by considering subleading corrections of the conformal block coming from a  heavy, time-dependent background state on the a point function of light operators, we evaluate  correlation function of four light operators.

\subsection{Conformal blocks at large central charge}

The monodromy method provides a way of computing conformal blocks  in the $c\to\infty$ semiclassical limit\footnote{An excellent review on this subject is given in  \cite{Fitzpatrick:2014vua} which we follow closely.}.  As was first conjectured in \cite{BELAVIN1984333,Hartman:2013mia} and later proven in \cite{Besken:2019jyw}, the leading order behavior of the conformal block for the $n$-point function for a set of primary operators $\{\co_{H_i}(x_i)\}$ with conformal weights $H_i$, under the assumption 
\be\label{ei}
 H_i/c=\text{fixed},\qquad  c\to\infty,
\ee
is given by
\begin{tcolorbox}[colback=yellow!7, colframe=blue!60!black]
\be\label{n-pt-monodromy}
\cf_{H}(x_i,h_p,H_i)\equiv\<\co_{H_1}(x_1)\dots\co_{H_n}(x_n)\>\simeq e^{-\tfrac{c}{6}f(x_i,h_p,H_i)}.
\ee
\end{tcolorbox}
Here, $f(x_i)$ is a function that needs to be specified and $h_p$ stands for the  conformal weights of  internal primaries exchanged. The monodromy method allows us to determine the function $f(x_i)$ in the semiclassical limit using the following trick. We consider the correlation function\footnote{One can show explicitly \cite{Fitzpatrick:2014vua} that since $\j$ is a light operator, in the large $c$ regime factorizes out of the correlator as the wavefunction $\j(z,x_i)$ and appears on the right hand side as in \eqref{Psi}. } 
\be\label{Psi}
\widetilde\cf_{H}(x_i,z)\equiv \<\co_{H_1}(x_1)\dots\hat\j(z)\dots\co_{H_n}(x_n)\>=\j(z,x_i)\cf(x_i),
\ee
where $\hat\j$ is the state-operator dual of a primary state with a null state at the second level of descendants, i.e it satisfies the condition 
\be\label{L-psi}
L_{\j}|\j\>\equiv\left(L_{-2}-\frac{3}{2(2h_\j+1)}L_{-1}^2\right)|\j\>=0.
\ee
We promote $L_{-n}$  acting on   $X\equiv\co_{H_1}(x_1)\dots \co_{H_n}(x_n)$, to differential operators \cite{DiFrancesco:1997nk}, through the relation
\be\label{Ln-differential}
\begin{split}
\<(L_{-n}\j(z))X\>&=\frac{1}{2\p i}\oint_w dz (z-w)^{1-n}\<T(z)\j(z)X\>\\
&=\sum_i\Big\{\frac{(n-1)H_i}{(w_i-w)^n}-\frac{1}{(w_i-w)^{n-1}}\del_{x_i}\Big\}\<\j(z)X\>.
\end{split}
\ee
Acting with $L_\j$ on \eqref{Psi}, using \eqref{Ln-differential} and substituting \eqref{n-pt-monodromy} in \eqref{Psi}   we obtain the second order differential equation\footnote{Here the  $\del_{x_i}\j$ terms are of order $\co(c^0)$, so  in the large $c$-limit they are suppressed.} 
\be\label{psi-ode}
\del_z^2\j(z)+\frac{6}{c}\,T(z)\j(z)=0,
\ee
where $T(z,x_i)$ is the expectation value of the energy momentum tensor given by,
\be\label{T-ode}
T(z)=\sum_{i=1}^n\frac{H_i}{(z-x_i)^2}-\frac{c_i}{z-x_i},\qquad c_i(x_i)\equiv\frac{c}{6}\frac{\del f(x_i)}{\del x_i}\,.
\ee
 Therefore, starting from \eqref{n-pt-monodromy} we have reduced the problem of finding $f(x_i)$, to the  solution of a differential equation which includes the derivatives of $f(x_i)$ via the {\it accessory parameters} $c_i$. Thus, the functions $f(x_i)$ are determined by integrating the $c_i(x_i)$ functions. 
 
 One can  introduce further constraints on the accessory parameters by employing the conformal  Ward identities, 
\begin{align}\label{conformal-WI}
\<T(z)X\>&=\sum_{i=1}^n
\left(\frac{H_i}{(z-x_i)^2}-\frac{1}{(z-x_i)}\del_i\right)
\<X\>=-\frac{c}{6}\,T(z,x_i)\<X\>.
\end{align}
Specializing to the case $n=4$, expanding \eqref{conformal-WI} for large $z$ i.e ${x_i}/{z}\to 0$  and imposing  $1/z^4$ behavior  in this limit, all terms up to third order should vanish. Thus, setting to zero the coefficients  of $z^{-3},z^{-2}$ and $z^{-1}$, we obtain respectively the following constraints 
\be\label{accessory-eq}
\sum_{i=1}^n c_i=0,\qquad \sum_{i=1}^n\left(c_i x_i-\frac{6h_i}{c}\right)=0,\qquad \sum_{i=1}^n\left(c_ix^2_i-\frac{12h_i}{c}x_i\right)=0\,.
\ee
Three constraints leave $(n-3)$ accessory parameters unfixed and thus as expected  $n=2$ and $n=3$ correlation functions are completely determined by conformal symmetry for all $c$. Furthermore, from a quick glance in \eqref{T-ode} we observe that the differential equation \eqref{psi-ode} has $n$ poles at $z=x_i$ and therefore the solutions are multivalued. Since we are interested in contributions only from  the identity block, common knowledge demands that solutions should be single valued around loops that contain pairs of operators that are operator product expansion (OPE)-contracted to the identity operator and imposing this condition we can determine the accessory parameters and obtain the expressions for $f(x_i)$.

\subsection{Subleading corrections and dressed two-point functions}

The previous discussion was limited to the leading order behavior of the conformal block. We now wish to extend our discussion by including  subleading correction to  \eqref{n-pt-monodromy}. This was first undertaken in \cite{Vos:2025voq} for two-point functions which  we now review  as a warm up for the four-point function that follows. 

The leading order behavior of the $n$-point function \eqref{n-pt-monodromy}  assumes that it contains a chain on heavy operators satisfying \eqref{ei}. Formally,  one way to consider subleading corrections   \eqref{n-pt-monodromy} is by introducing operators with light conformal weights that can be used as perturbations. These belong to the third class of operators introduced in \eqref{operator-classes}. In this case the correlation function describes the behavior of a set of    light probes in a heavy background and is given by  
\be
\label{FHh}
\cf^{(n,m)}_{H,h}(x_i,z_j)=\<\co_{H_1}(x_1)\dots\co_{H_n}(x_n)\co_{h_1}(z_1)\dots\co_{h_m}(z_m)\>,
\ee
where  $H_i$ denote the heavy and $h_j$ the light hermitian operators and  as indicated by \eqref{operator-classes} satisfy 
\be\label{H-h-definition}
\frac{h_j}{c}=\co(1/c),\qquad \frac{H_i}{c}=\text{fixed}\,.
\ee
The leading term is not  affected by the presence of light operators at $c\gg1$ limit and therefore such contributions should be separately captured  by  subleading  corrections that scale as $\co(c^0)$. Therefore,   the conformal block \eqref{FHh} under the assumptions \eqref{H-h-definition} factorizes as follows,
\begin{tcolorbox}[colback=yellow!7, colframe=blue!60!black]
\be\label{Fnm}
\cf^{(n,m)}_{H,h}(x_i,z_j)=e^{-\frac{c}{6}f_0(H;x_i)}e^{f_1(H,h;x_i,z_j)}\,\,,
\ee 
\end{tcolorbox}
where $f_1(H,h;x_i,z_j)$ arises from the heavy-light interactions. 

The problem of finding the functions $f_0$ and $f_1$ is reduced to the evaluation of the accessory parameters by  repeating the  trick discussed earlier i.e by considering 
\be
\widetilde \cf^{(n,m)}_{H,h}=\<\co_{H_1}(x_1)\dots\co_{H_n}(x_n)\hat\j(z)\co_{h_1}(z_1)\dots\co_{h_m}(z_m)\>=\j(z;x_i,z_j)\cf^{(n,m)}_{H,h}(x_i,z_j),
\ee
with $\j$ a  degenerate state satisfying \eqref{L-psi}. Following the steps of the previous section one can easily observe that the wavefunction  $\j(z;x_i,z_j)$ satisfies the differential equation,
\be\label{psi-x-y}
\j''(z;x_i,z_j)+\frac{6}{c}\,T(z,x_i,z_j)\j(z;x_i,z_j)=0,
\ee
where $f'(z)\equiv\frac{d}{dz}f(z)$ and the energy momentum tensor now reads,
\be\label{Tz-dressed}
T(z;x_i,z_j)=\sum_{i=1}^n\left(\frac{H_i}{(z-x_i)^2}-\frac{C^{(0)}_{i}}{z-x_i}+\frac{C^{(1)}_{i}}{z-x_i}\right)+\sum_{i=1}^m\left(\frac{h_i}{(z-z_i)^2}+\frac{\tilde c_{i}}{z-z_i}\right).
\ee
The accessory parameters are given by
\be
C^{(0)}_{i}\equiv\frac{c}{6}\frac{\del f_0(x_i)}{\del x_i},\qquad C^{(1)}_{i}\equiv\,\frac{\del f_1(x_i,y_j)}{\del x_i}\,,\qquad \tilde c_{j}\equiv\,\frac{\del f_1(x_i,y_j)}{\del y_j}\,.
\ee
Returning to \eqref{Tz-dressed} we can separate between heavy and light sector\footnote{From now on we neglect the $(x_i,z_j)$ argument for simplicity and restore it whenever it is necessary.},
\begin{align}
T_H(z)&\equiv\sum_{i=1}^n\left(\frac{H_i}{(z-x_i)^2}-\frac{C^{(0)}_{i}}{z-x_i}\right)\,,\\
T_h(z)&\equiv\sum_{i=1}^m\left(\frac{h_i}{(z-z_i)^2}+\frac{\tilde c_{i}}{z-z_i}\right)+\sum_{i=1}^n\frac{C^{(1)}_{i}}{z-x_i}\,,
\end{align}
and we can use $T_h$ as a perturbation of $T_H$.  As a second order differential equation \eqref{psi-x-y} has two solutions denoted as $\j_\pm(z)$ and we  split them up into a leading part $\rho_{\pm}(z)$ plus a subleading correction $\eta_{\pm}(z)$  
\be\label{psi-expansion}
\j_\pm(z)=\r_\pm(z)+\eta_\pm(z)\,,
\ee
satifying separately the differential equations, 
\begin{tcolorbox}[colback=yellow!7, colframe=blue!60!black]
\be\label{r-eta}
\r''_\pm(z)+\frac6c\, T_H(z)\r_\pm(z)=0\,,\qquad\quad\eta''_\pm(z)+\frac6c\,T_H(z)\eta_\pm(z)=-\frac{6}{c}\,T_h\r_\pm(z)\,.
\ee
\vskip0.0001cm
\end{tcolorbox}
 The $\eta_\pm(z)$ functions satisfy a subleading inhomogenous version of the leading homogeneous equation satisfied by $\r_\pm(z)$. 
 We will assume that the leading basis of solutions is Wronskian normalized
 \begin{equation}\label{Wronskian}
     \rho_+(z)\rho_-'(z) - \rho_+'(z)\rho_-(z) =1.
 \end{equation}
 As it was shown in \cite{Fitzpatrick:2014vua,Vos:2025voq} one can use variation of parameters to express $\eta_\pm$ in terms of $\r_\pm$ in the form, 
\be\label{eta-form}
\eta_\pm(z)=A_\pm(z)\r_+(z)+B_\pm(z)\r_-(z),
\ee
with known derivatives for $A_\pm$ and $B_\pm$ given by,
\be\label{Apm-Bpm}
A'_\pm(z)=-T_h(z)\r_\pm(z)\r_+(z),\qquad \quad B'_\pm(z)=T_h(z)\r_\pm(z)\r_-(z)\,.
\ee
Both equations in \eqref{r-eta} contain poles which imply that  solutions $\r_\pm(z)$ and $\eta_\pm(z)$ are in principle multivalued since they depend on whether the curve of integration includes poles or not. However, considering contributions only from the identity block the solutions become  single valued, so we avoid such  complications. Our main focus is to express $\eta_\pm(z)$ in terms of $\r_\pm(z)$ which in turn cannot be obtained in a closed form. Considering a curve $\cc$ including the poles $z_0=(x_i,y_j)$, the $\eta_\pm(z)$ solutions are shifted by,  
\be\label{eta-monodromy}
\eta_\pm(z)\to \eta_\pm(z) + \left[\oint_\cc dw\,A'_\pm(w)\right]\r_+(z)+\left[\oint_\cc dw\, B'_\pm(w)\right]\r_-(z)\,,
\ee
under the assumption  that $\r_\pm(z)$ are single valued at leading order. 

\subsection*{Two-point functions} 

Now keeping the number of heavy operators general in \eqref{FHh}, we consider two light operators in the heavy background, namely 
\be\label{F-2n}
\cf_{H,h}^{(2n,2)}=\<\co_{H_1}(x_1)\dots\co_{H_n}(x_n)\co_{h}(z_2)\co_{h}(z_1)\co_{H_{n}}(x_{n})\dots\co_{H_{1}}(x_{1})\>,
\ee
and by employing  the heavy state definition $
|\mathcal H\>=\co_{H_1}(x_1)\dots\co_{H_n}(x_n)|0\>$ given in  \eqref{Heavy-states},
then \eqref{F-2n} acquires the  form,   
\be
\cf_{H,h}^{(2n,2)}=\<\mathcal H|\co_{h}(z_2)\co_{h}(z_1)|\mathcal H\>,
\ee
This corresponds to a two-point function of  two light operators in an arbitrary heavy background. Considering now the general equations \eqref{r-eta} in our set up and taking  contributions only from the identity block by imposing single valued constraints in \eqref{eta-monodromy} integrals,  we obtain\footnote{Details of the computations are given in \cite{Vos:2025voq}.}
\be
f_1(z_1,z_2)=2h\log\left(\r_+(z_1)\r_-(z_2)-z_1\leftrightarrow z_2\right)
\ee
which leads directly to the subleading correction of the two-point function for light operators, 
\begin{tcolorbox}[colback=yellow!7, colframe=blue!60!black]
\be\label{Q-2pt}
Q(z_1,z_2)\equiv e^{f(z_1,z_2)}=\frac{1}{(\r_+(z_1)\r_-(z_2)-\r_+(z_2)\r_-(z_1))^{2h}}\,.
\ee
\end{tcolorbox}
The  singular behavior at $z_1\to z_2$ structure follows the general behavior  of the CFT two-point functions with the difference that the denominator is dressed by the solutions of the heavy operator functions $\r_\pm(z)$.

\subsection{Dressed four-point functions}

\begin{figure}
	\centering
		\includegraphics[scale=0.8]{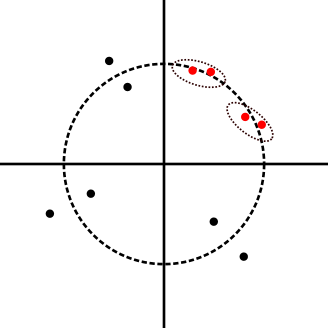}
	\caption{Four light operators (red dots) are fused together pairwise by the OPE to the identity operator. The black dots represent the heavy operators of both the state $|\ch\rangle$ and its adjoint state $\langle \ch|$.}
	\label{FourPointMonodromy}
\end{figure}

Let us now move on to the case of four light operators.  To compute the conformal identity block of the light sector, we will divide the light operators up into two pairs and contract them pairwise using their OPE relations. Then  we will truncate the OPE by only summing over descendants of the identity operator as in Figure \ref{FourPointMonodromy}. In this case, we have two sets of contractions and as a result two cycles that need to have trivial monodromy. Applying the same method with the two-point function, the poles at the light operator locations only manifest themselves at subleading order. In other words, the leading basis of solutions $\rho_{\pm}(z)$ is agnostic to the presence of the light operators at $z_3$ and $z_4$. This in turn implies that within any given cycle, the only information that enters from the light sector comes from the residues of the simple poles contained within the individual cycles.
By performing the method explained above we expand the full solutions $\psi_{\pm}(z)$  as in \eqref{psi-expansion}
where now $\eta_\pm(z)$ is given by \eqref{eta-form}. Then we find two sets consisting of each three independent constraints,
\begin{align}
\textbf{Cycle 1:}\qquad& \sum_{i=1,2}\left(h\,\big(\rho_{\pm}(z_i)\rho_{+}(z_i)\big)' + c_1\,\rho_{\pm}(z_i)\rho_+(z_i)\right) =0\,,\nonumber \\
&\sum_{i=1,2}\left(h\,\Big(\rho_{\pm}(z_i)\rho_{-}(z_i)\Big)'+ c_2\,\rho_{\pm}(z_i)\rho_-(z_i)\right)=0\,,
\end{align}
and
\begin{align}
\nonumber\textbf{Cycle 2:}\qquad& \sum_{i=3,4}\left(h\,\big(\rho_{\pm}(z_i)\rho_{+}(z_i)\big)' + c_1\,\rho_{\pm}(z_i)\rho_+(z_i)\right) =0\,, \\
&\sum_{i=3,4}\left(h\,\Big(\rho_{\pm}(z_i)\rho_{-}(z_i)\Big)'+ c_2\,\rho_{\pm}(z_i)\rho_-(z_i)\right)=0\,.
\end{align}
We can view  them as a system of partial differential equations for the function $f_1(z_1,z_2,z_3,z_4)$ which enters the constraints through the accessory parameters
\begin{equation}
    c_i = \partial_{z_i}f_1(x_i).
\end{equation}
Since the cycle 1 constraints only depend on $z_1$ and $z_2$ and the cycle 2 constraints only depend on $z_3$ and $z_4$ and since the dependence on $f_1(z_i)$ only enters through its derivatives there is an obvious solution to these constraints given simply  by the sum of two solutions,
\begin{align}
    & f_1(z_1,z_2,z_3,z_4) = \,-2h\log \left(\rho_+(z_1)\rho_-(z_2) - z_1\leftrightarrow z_2\right) \nonumber \\
    &\qquad\qquad\qquad\qquad - 2h\log \left(\rho_+(z_3)\rho_-(z_4) - z_3\leftrightarrow z_4\right).
\end{align}
Therefore, defining the four-point function as 
\be
Q(z_1,z_2,z_3,z_4)=e^{f_1(z_1,z_2,z_3,z_4)}\,,
\ee
leads to the following factorized version for the light sector contribution to the identity block,
\begin{tcolorbox}[colback=yellow!7, colframe=blue!60!black]
  \begin{equation}
     Q(z_1,z_2,z_3,z_4) =  \frac{1}{\left(\rho_+(z_1)\rho_-(z_2) - z_1\leftrightarrow z_2\right)^{2h}}\frac{1}{\left(\rho_+(z_3)\rho_-(z_4) -  z_3\leftrightarrow z_4\right)^{2h}}.
\end{equation}  
\end{tcolorbox}
The leading factorized nature of the expectation value of four light operators indicates  that this level of approximation is not detailed enough. In particular, to learn  how perturbation by one operator in the past affects the measurement of another observable in the future we require some level of coupling. In such a case one may proceed, either directly by coupling the light operators or indirectly by letting the operators backreact with the heat bath. We will apply the latter option in the next section. 

The arguments for the four-point function can be easily generalized to the 2$n$-point function of light operators giving
\be
Q(z_1,\dots,z_{2n})=\prod_{i=0}^{n-1}\frac{1}{\big(\rho_+(z_{2i+1})\rho_-(z_{2i+2}) - z_{2i+1}\leftrightarrow z_{2i+2}\big)^{2h}}\,\,.
\ee
The reason is that  the pattern above repeats for a greater number of pairs of light operators. Every loop that contains a pair of OPE-contracted light operators introduces a set of disconnected constraint equations. Solving the constraint equations for each loop introduces a new term in $f_1(z_i)$.

\section{Quantum chaos from time-dependent states}

\label{sec:quantumchaos}

In this section we now exploit the tools presented in previous sections to extract chaos from the four-point function consisting of two middle-heavy and two light operators. In particular we consider the operator ordering between two middle-heavy and two light operators and show how to treat the arising discontinuities. Since under general principles \cite{Maldacena:2015waa} the alternating terms, for chaotic systems should be of the form
\begin{equation}\label{Lyapunov}
    \mathcal F^{(2n,4)}_{H,h}(t) = \langle \co_L\co_L\rangle \langle \co_M\co_M\rangle \left(1- \frac{u}{c}e^{\lambda_L t}\right),
\end{equation}
for $h=(h_M,h_L)$ and $u$ some constant that will be specified below, by turning on the Lorentzian time we evaluate the Lyapunov exponent which saturates the MSS bound.

\subsection{Middle-heavy operators}

As was discussed in the previous section, if we take the light operator sector of the identity block contribution to the four light operator expectation value,
\begin{equation}
    \langle \mathcal H| \wick{ \c1 \cO_L(z_1) \c1 \cO_L(z_2) \c2 \cO_L(z_3) \c2 \cO_L(z_4) } |\mathcal H\rangle,
\end{equation}
with OPE contractions  to leading order in the large-$c$ limit, the expectation value just factorizes in the following way,
\begin{equation}
    \langle \mathcal H| \wick{ \c1 \cO_L(z_1) \c1 \cO_L(z_2) \c2 \cO_L(z_3) \c2 \cO_L(z_4) } |\mathcal H\rangle = \langle \mathcal H|  \wick{ \c \cO_L(z_1) \c \cO_L(z_2)}    |\mathcal H\rangle \langle \mathcal H|  \wick{ \c \cO_L(z_3) \c \cO_L(z_4)}    |\mathcal H\rangle.
\end{equation}
This suggests that the leading order in the light sector is not sufficient to observe the effects of quantum chaos. In order to witness chaotic effects we have to include a way to systematically add subleading corrections at large-$c$. We will use a method inspired by \cite{Roberts:2014ifa}, we will promote one of the pairs of light operators to a pair of middle-heavy operators as described in \eqref{operator-classes}. 
We will consider the presence of the middle-heavy operators as a perturbation of the heavy state $|\mathcal H\rangle$, namely $|\mathcal{H}\rangle \to |\mathcal{\tilde{H}}\rangle$ with $|\mathcal{\tilde{H}}\>$ given in \eqref{H-tilde}. The new state has a modified normalized stress tensor expectation value
\begin{equation}
    T_M(z) \equiv \frac{\langle \tilde{\mathcal{H}}|T(z)|\tilde{\mathcal{H}}\rangle}{\langle \tilde{\mathcal{H}}|\tilde{\mathcal{H}}\rangle},
\end{equation}
and is given by 
\begin{align}
     \frac{6}{c}T_M(z) &= \frac{6}{c}T_H(z) + \delta T(z) \nonumber \\
    &=\frac{6}{c}T_H(z) + \sum_M \frac{6 \zeta}{(z-w_j)^2} +\frac{c}{6}\frac{c_j}{z-w_j} + \sum_H \frac{6}{c}\,\frac{\delta c_i}{z-x_i}.  
\end{align}
We included the shift in the heavy sector accessory parameters $\delta c_i$ coming from the introduction of the $\co_M$ operators. We will make the conservative assumption that $\frac{6}{c}\delta c_i \propto \zeta$.
The presence of the extra terms in $T_M(z)$ compared to $T_H(z)$ modifies the solutions of the linear Fuchsian equation
\begin{equation}
    \left(\rho^{M}_{\pm}\right)'' + \frac{6}{c} T_M(z) \rho^{M}_{\pm}(z) = 0.
\label{modifiedFuchsian}
\end{equation}
We can consider the extra terms in $T_M(z)$ compared to $T_H(z)$ as a small perturbation. Repeating the steps of section \ref{Conformal-blocks} by considering the heavy operator to be $\co_M$, we expand the corresponding $\r_\pm$ functions as follows \\
\begin{equation}\label{rho-M}
    \rho_{\pm}^M (z)= \rho_{\pm}(z) + \eta^M_{\pm}(z).
\end{equation}
Substituting \eqref{rho-M} in \eqref{modifiedFuchsian} we obtain the  equation for the first subleading correction\footnote{In contrast to \eqref{r-eta}, we absorb $(6/c)$-factor in the definition of $\d T(z)$.}
\begin{tcolorbox}[colback=yellow!7, colframe=blue!60!black]
\begin{equation}
    (\eta_\pm^M)''(z) + \frac{6}{c}\,T_H(z)\eta^M_\pm(z) = -\delta T(z)\rho_{\pm}(z).
\end{equation}
\end{tcolorbox}
This is an inhomogeneous version of the leading differential equations, hence $\eta(z)$ can be expressed in terms of the leading solutions by means of variation of parameters\footnote{Note that the expressions $A,B$ in principle differ from the ones in \eqref{eta-form} but we keep the same notation since they follow from the same technique.}
\begin{equation}
    \eta_{\pm}(z) = \rho_{\pm}(z) + A_{\pm}(z)\rho_+(z) + B_{\pm}(z)\rho_-(z)\,,
\end{equation}
with coefficient functions $A^M_\pm(z), B^M_{\pm}(z)$ defined implicitly through the first-order differential equations as in \eqref{Apm-Bpm}, namely
\begin{align}\label{coefficientfunctions1}
\frac{d}{dz} A_{\pm}(z) =  -\delta T(z)\rho_{\pm}(z)\rho_+(z), \qquad \frac{d}{dz} B_{\pm}(z) = \delta T(z) \rho_{\pm}(z)\rho_-(z). 
\end{align}
We can now return to the light sector of the identity block and  include the back-reaction of $\co_M$ to the heavy state and couple the light sector to the back-reacted state, reading
\begin{equation}
    Q_L(z_1,z_2) = \frac{\langle \mathcal{\tilde{H}}|\co_L(z_2)\co_L(z_1)|\mathcal{\tilde{H}}\rangle}{\langle \mathcal{\tilde{H}}|\mathcal{\tilde{H}}\rangle}.
\end{equation}
This is simply done by promoting the leading solutions of the Fuchsian to the corrected solutions,
\begin{equation}
    \rho_{\pm}(z) \rightarrow \rho^M_{\pm}(z) = \rho_{\pm} + A_{\pm}\rho_+ + B_{\pm}\rho_-\,.
\label{correctedsolutions}
\end{equation}
This results in the correlation function of two light operators \eqref{Q-2pt} in the presence of a backreacted background,  reading
\begin{tcolorbox}[colback=yellow!7, colframe=blue!60!black]
\begin{equation}
    Q_L(z_1,z_2) = \frac{1}{\left(\rho^M_+(z_1)\rho^M_-(z_2)-\rho^M_+(z_2)\rho^M_-(z_1)\right)^{2h_L}}.
\label{correctedlightsector}
\end{equation}
\end{tcolorbox}

\subsection{Operator ordering in the light and middle-heavy sector}

\begin{figure}
	\centering
		\includegraphics[scale=1]{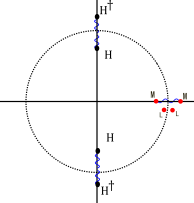}
	\caption{On the Euclidean plane the operators are radially ordered corresponding to the expectation value \eqref{principal-correlator}. As an example, a state $|\mathcal H\rangle$ consisting of two heavy operators has been drawn with a branch cut between the two middle-heavy operators. }
	\label{startingconfiguration}
\end{figure}

Having obtained the Euclidean expression for the backreacted light sector the next step is to consider the analytic continuation to Lorentzian time. In particular, we will first show how to internally swap the light and middle-heavy operators and subsequently turn on the time-evolution of the middle-heavy operators. We will proceed step-by-step and first consider how to swap the order of the operators. Recall that we aim to compute the normalized Lorentzian commutator,
\begin{equation}
    {C}_\ch(t)= - \frac{\langle \mathcal{H}|[\co_M(t),\co_L(0)]^2|\mathcal{H}\rangle}{\langle \mathcal{H}|\mathcal{H}\rangle}\,,
\end{equation}
and in accordance  with \cite{Maldacena:2015waa}, we will focus on the alternating terms
\begin{equation}
    \langle \mathcal{H}|\mathcal{H}\rangle \; \mathcal F^{(2n,4)}_{H,h}(t) \equiv -\langle \mathcal H|\co_L\co_M\co_L\co_M|\mathcal H\rangle -\langle \mathcal H|\co_M\co_L\co_M\co_L|\mathcal H\rangle.
\label{alternatingterms}
\end{equation}
The starting point is the expression for the Euclidean light sector
\begin{equation}
    Q_L(z_1,z_2) = \frac{\langle \mathcal{\tilde{H}}|\co_L(z_1)\co_L(z_2)|\mathcal{\tilde{H}}\rangle}{\langle \mathcal{\tilde{H}}|\mathcal{\tilde{H}}\rangle},
\end{equation}
which naturally corresponds to the operator ordering \eqref{principal-correlator}. To obtain  the ordering of the terms in \eqref{alternatingterms}, we  employ the prescription presented in section \ref{sec:LorentzianCFT}. The initial sheet operator positions are drawn in \ref{startingconfiguration}. Switching the operator ordering of two operators corresponds to forcing one operator to cross a branch cut associated with the second operator.

 We displayed the branch cut between the middle-heavy operator explicitly. We can compute the branch discontinuity across this cut. Consider the form of the corrected solutions $\rho_{\pm}^M(z)$ in \eqref{correctedsolutions}. The leading solutions $\rho_{\pm}(z)$ are agnostic to the positions of the middle-heavy operators and therefore should be regular at their insertion locations. Therefore the only source for the branch discontinuity can come from the coefficient functions $A_{\pm}(z), B_{\pm}(z)$. Since the derivatives of the coefficient functions are given by \eqref{coefficientfunctions1}, we can compute the branch discontinuity by contour integration across the cut. Take $\mathcal{C}$ to be a closed contour that starts at a point $z$, crosses the cut and then returns onto itself. Under this curve the coefficients pick up the following branch discontinuities
\be\label{discontinuity1}
 A_{\pm}\rightarrow  A_{\pm} + \oint_{\mathcal C} dz  A'_{\pm}(z),\qquad B_{\pm}\rightarrow  B_{\pm} + \oint_{\mathcal C} dz B'_{\pm}(z)\,. 
\ee
 Recall that the difference between the values at the initial and the final point of a contour is simply given by the integral of the derivative of the function along the contour.

The derivatives of the coefficient functions are given in \eqref{coefficientfunctions1}, so we can evaluate the contour integrals. The function $\delta T(z)$ has poles at the locations of the middle-heavy operators so the exact discontinuities depend on the details of the contour. 

Before we consider the individual terms there is one more component we require. The residues of $\delta T(z)$ depend explicitly on the accessory parameters $c_j(w_i)$ of the middle-heavy operators. We can exploit the Virasoro Ward identity in order to obtain these accessory parameters to leading order. Note that these parameters are defined as the derivatives of the logarithm of the conformal block. To leading order the middle-heavy sector decouples from the light sector and  hence is also given by the factorized expression
\begin{equation}
    Q_M(w_1,w_2) = \frac{1}{\left(\rho_+(w_1)\rho_-(w_2)-\rho_+(w_2)\rho_-(w_1)\right)^{2h}}\,.
\end{equation}
Defining the quantities,
\begin{equation}
    Q_{M}(w_i) = e^{f_M^{(1)}(w_i)}, \qquad\quad c_i(w_i) = \frac{\partial}{\partial_{w_i}} f^{(1)}_{M}(w_i),
\end{equation}
we can explicitly express the accessory parameters in terms of the leading order solutions
\begin{align}
    & c_1(w_i) = -2h_M  \frac{\rho_+'(w_1)\rho_-(w_2)-\rho_+(w_2)\rho_-'(w_1)}{\rho_+(w_1)\rho_-(w_2)-\rho_+(w_2)\rho_-(w_1)}, \label{middleaccessoryparameters1}\\
    & c_2(w_i) = -2h_M  \frac{\rho_+(w_1)\rho_-'(w_2)-\rho_-(w_1)\rho_+'(w_2)}{\rho_+(w_1)\rho_-(w_2)-\rho_+(w_2)\rho_-(w_1)}. \label{middleaccessoryparameters2}
\end{align}

\subsubsection*{Alternating terms}

We  now proceed to evaluate the two individual alternatingly-ordered terms in \eqref{alternatingterms}. Recall that the two terms that appear in \eqref{alternatingterms} are given by,
\begin{equation}
    \langle \mathcal H| \cO_L \cO_M \cO_L \cO_M |\mathcal H\rangle, \qquad \langle \mathcal H| \cO_M \cO_L \cO_M \cO_L |\mathcal H\rangle. 
\label{alternatingterms2}
\end{equation}
neglecting We will describe the procedure for both terms in turn. The first term in \eqref{alternatingterms2} is obtained through the operator switching,
\begin{equation}
   \langle \mathcal H| \overset{\scalebox{2}{$\curvearrowleft$}}{\cO_M \cO_L} \cO_L \cO_M |\mathcal H\rangle\longrightarrow \langle \mathcal H| \cO_L \cO_M \cO_L \cO_M |\mathcal H\rangle.
\end{equation}
As we have already pointed out, a switch of the light operator at $z_2$ is accompanied by a  crossing of  the branch cut associated with the middle-heavy operator located at $w_2$. This is achieved by taking the light operator located at $z_2$ and moving it a full clockwise circle around the middle-heavy operator located at $w_2$, see Figure \ref{orderingfigure}, we will call this contour $\mathcal{C}_2$. Similarly  the second correlator  in \eqref{alternatingterms2} is obtained by performing the following  swap
\begin{equation}
   \langle \mathcal H| \cO_M \cO_L \overset{\scalebox{2}{$\curvearrowright$}}{\cO_L \cO_M} |\mathcal H\rangle\longrightarrow \langle \mathcal H| \cO_M \cO_L \cO_M \cO_L |\mathcal H\rangle.
\end{equation}
This can be implemented by making the operator located at $z_1$ perform a counter-clockwise loop around the operator located at $w_1$, see Figure \ref{orderingfigure}, we denote this contour by $\mathcal C_1$.

In both cases we can compute the branch discontinuities. Take the contour $\mathcal{C}_2$, only the location $z_2$ moves across the branch point and the functions evaluated at $z_1$ stay invariant
\begin{align}\label{rho-z1z2}
  \nonumber  & \rho_{\pm}^M(z_1) \rightarrow \rho_{\pm}^M(z_1)\\
    & \rho_{\pm}^M(z_2) \rightarrow \rho_{\pm}(z_2) + \left(A_{\pm}+\oint_{\mathcal{C}_2}dz A'_{\pm}\right)\rho_{+}(z_2) + \left(B_{\pm}+\oint_{\mathcal{C}_2}dz B'_{\pm}\right)\rho_{-}(z_2).
\end{align}
 The discontinuities are given by inserting \eqref{coefficientfunctions1} into the contour integrals \eqref{discontinuity1}, and realizing that within the contour $\rho_{\pm}(z)$ are regular. We can thus perform the contour integration using Cauchy's theorem by computing the residues of the poles of the stress tensor expectation value 
\begin{align}
    & \frac{1}{2\pi i} \oint_{\mathcal C_2}dz A'_{\pm}= \frac{6}{c}\, c_2 \rho_{\pm}(w_2)\rho_+(w_2) + 6 \zeta \left(\rho_{\pm}'(w_2)\rho_+(w_2) + \rho_{\pm}(w_2)\rho_+'(w_2)\right), \label{c2Adiscontinuity}\\
    & \frac{1}{2\pi i} \oint_{\mathcal C_2}dz B'_{\pm}= -\frac{6}{c}\, c_2 \rho_{\pm}(w_2)\rho_-(w_2) - 6\zeta \left(\rho_{\pm}'(w_2)\rho_-(w_2) + \rho_{\pm}(w_2)\rho_-'(w_2)\right). \label{c2Bdiscontinuity}
\end{align}

Similarly, we can apply the $\mathcal{C}_1$ contour integration to  the $\cO_M\cO_L\cO_M\cO_L$-correlator. In this case all the functions inside \eqref{correctedlightsector} evaluated at $z_1$ stay invariant and those evaluated at $z_2$ pick up the following discontinuities,
\begin{align}\label{rho-z2z1}
   \nonumber & \rho_{\pm}^M(z_1) \rightarrow \rho_{\pm}(z_1) + \left(A_{\pm}+\oint_{\mathcal{C}_1} dz A'_{\pm}\right)\rho_{+}(z_1) + \left(B_{\pm}+\oint_{\mathcal{C}_1}dz  B'_{\pm}\right)\rho_{-}(z_1) \,,\\
    & \rho_{\pm}^M(z_2) \rightarrow \rho_{\pm}^M(z_2).
\end{align}
Through the same method, by taking into account the sign change due to the orientation of the contour $\mathcal{C}_1$, we now obtain
\begin{align}
    & \frac{1}{2\pi i} \oint_{\mathcal C_1}dz  A'_{\pm}= -\frac{6}{c}\, c_1 \rho_{\pm}(w_1)\rho_+(w_1) - 6 \zeta \left(\rho_{\pm}'(w_1)\rho_+(w_1) + \rho_{\pm}(w_1)\rho_+'(w_1)\right), \label{c1Adiscontinuity} \\
    & \frac{1}{2\pi i} \oint_{\mathcal C_1} dz B'_{\pm}= \frac{6}{c} c_1 \rho_{\pm}(w_1)\rho_-(w_1) + 6\zeta \left(\rho_{\pm}'(w_1)\rho_-(w_1) + \rho_{\pm}(w_1)\rho_-'(w_1)\right) \,.\label{c1Bdiscontinuity}
\end{align}
This completes the prescription of the terms associated with the alternating orderings.

\subsection{Turning on Lorentzian time}

We now encounter the second subtlety regarding the evolution to Lorentzian time.  Since we have a prescription that gives us the desired ordering, we now need a trajectory on the Lorentzian cylinder corresponding to turning on Lorentzian time. The middle-heavy operators are evaluated at a time $t$ and thus in the presence of a time-dependent background state $|\mathcal H\rangle$ this requires  more attention. 

The middle-heavy operator $\cO_M(t)$ has to be moved from an initial time 0 to a final time $t$. Imagine that the trajectory on the two-complex dimensional plane is parametrized by a parameter $s\in[0,t]$. As described in section \ref{sec:LorentzianCFT}, there are many trajectories on the $(z,\bar{z})$ plane that start and end at the given points but there is a particularly convenient parametrization that is given by $z(s) = e^{2is}z_0$ and $\bar{z}(s)=\bar{z}_0$.
A straightforward way of implementing the parametrization is by performing a transformation of the leading Fuchsian ODE, namely
\begin{equation}\label{F-ODE}
    \rho_{\pm}''(z) + \frac{6}{c}\,T(z)\rho_{\pm}(z)=0. 
\end{equation}
As demonstrated in appendix A of \cite{Vos:2025voq}, if $\rho_{\pm}(z)$ solves \eqref{F-ODE}, then under the substitution $z\rightarrow u(z)$ the following ODE is solved,
\begin{equation}
    \tilde{\rho}_{\pm}(u)'' + \frac{6}{c}\, \tilde{T}(u)\tilde{\rho}_{\pm}(u) = 0.
\label{secondODE}
\end{equation}
where the individual objects are given by
\begin{equation}
    \tilde{\rho}_{\pm} = \frac{1}{\sqrt{z'(u)}} \rho_{\pm}(u),\qquad \frac{6}{c} \tilde{T}(u) = \frac{6}{c}\left(z'(u)^2 T(z(u)) + \frac{c}{12} \{z,u\}\right)\,.
\end{equation}
By taking $s=u(z) = \log(z)/(2i)$ we obtain the following\footnote{An extra subtlety worth  noting is that this transformation should not be interpreted as a conformal transformation. We are parametrizing a curve on the two-complex dimensional $(z,\bar{z})$-plane. Note that we could have performed a conformal transformation instead. In that case the operators inside the correlator pick up transformation factors as well. While in either case the physical interpretation changes slightly, the final conclusion is not altered by either choice.}
\begin{equation}
    \frac{6}{c}\tilde{T}(s) =1 -\frac{24}{c} e^{4is}T(e^{2is})\,.
\end{equation}
The remarkable property, as was shown in \cite{Vos:2025voq}, is that for real-valued $s$,  $\frac{6}{c}\tilde{T}(s)$ is real-valued. This can be shown from the Hermitian adjoint rule of the stress tensor operator $\hat{T}(z)^{\dagger}=(1/z^*)^4\hat{T}(1/z^*)$. The transformed ODE is given by
 \begin{equation}\label{ODE2}
     \tilde{\rho}_{\pm}(s)'' + \left(1-e^{4is}T(e^{2is})\right)\tilde{\rho}_{\pm}(s)=0\,.
 \end{equation}
Equation \eqref{ODE2} has a potential that first is real-valued and second periodic  under $s\rightarrow s+\pi$ which implies that it  is a Hill type equation. We can invoke Floquet's theorem and claim that there exists a basis of solutions that factorizes into an exponential part and a periodic part
\begin{equation}
    \tilde{\rho}_{\pm}(s) = e^{\pm i\alpha s} p_{\pm}(s),
\end{equation}
with $p_{\pm}(s+\pi) = p_{\pm}(s)$. We can therefore express the reparametrized solutions as
\begin{equation}
    \rho_{\pm}(e^{2is}) = \sqrt{2i} e^{is} e^{\pm i\alpha s}p_{\pm}(s).
\end{equation}
There are two relevant cases, either $\alpha$ is purely real or purely imaginary \cite{MagnusWinkler}. In \cite{Vos:2025voq} it was shown that the `unstable' purely imaginary case leads to thermal expectation values in the late time-limit in the case of the two-point expectation value. Following that conclusion, we will assume that our initial state $|\mathcal{H}\rangle$ is such that its associated Floquet exponent is purely imaginary, namely
\begin{equation}
    \rho_{\pm}(e^{2is}) = \sqrt{2i}  e^{(\mp |\alpha| +i)s}p_{\pm}(s).
\end{equation}
Therefore we will also assume that $|\mathcal H\rangle$ is such that we can substitute the above expression into $\rho_{\pm}(w_i)$, by keeping in mind that the points $w_i$ are the only ones that undergo the time-evolution trajectory. The branch discontinuity terms in the subleading correction part of $\rho^M_{\pm}(z)$ are the only terms that carry explicit dependence on $w_i$. Therefore we only have to consider the effect of the reparametrization on the branch discontinuity terms \eqref{c2Adiscontinuity}, \eqref{c2Bdiscontinuity},    \eqref{c1Adiscontinuity} and \eqref{c1Bdiscontinuity}. We refer the reader to appendix \ref{sec:realtimeappendix} for the detailed expressions. The important fact though  is that after reparametrization the branch discontinuities also factorize into a product of an exponential factor and a bounded factor respectively,
\begin{align}\label{AB-behavior}
    & \oint_{\mathcal C_i}dz  A'_{\pm} \equiv \zeta e^{(-|\alpha|\mp|\alpha|)t} \mathcal{A}^{(i)}_{\pm}(t), \qquad \oint_{\mathcal C_i} dz B'_{\pm} \equiv \zeta e^{(|\alpha|\mp|\alpha|)t} \mathcal{B}^{(i)}_{\pm}(t),
\end{align}
where the bounded factors $\mathcal A^{(i)}_{\pm}, \mathcal B^{(i)}_{\pm}$ satisfy the following constraints 
\begin{align}\label{AB-constraints}
    & \mathcal{A}^{(1)}_{\pm}(t) = \mathcal{A}^{(2)}_{\pm}(t), \qquad \mathcal{B}^{(1)}_{\pm}(t) = \mathcal{B}^{(2)}_{\pm}(t).
\end{align}
Due to the pattern of enhancements and cancellations in the exponents we can see that the branch discontinuities fall into three classes based on their behavior at large time
\begin{align}\label{disc-behavior}
    \nonumber \text{{\bf growing}\,\,:} \qquad &\oint_{\mathcal C_i}dz B'_-, \\
     \text{{\bf stable}\,\,:} \qquad &\oint_{\mathcal C_i}dz A'_-\,,\, \,\oint_{\mathcal C_i}dz B'_+\,, \\
   \nonumber \text{{\bf decaying}\,\,:} \qquad &\oint_{\mathcal C_i}dz A'_+\,. 
\end{align}
Going back to  the two-point function of the light sector on the principal sheet \eqref{correctedlightsector},
we  obtain the $\cO_M \cO_L\cO_M\cO_L$-ordering by replacing  \eqref{AB-behavior} in \eqref{rho-z1z2} and \eqref{rho-z2z1}, giving
\begin{align}
   \nonumber &\rho_{+}^M(z_1)\rightarrow \rho_+(z_1) + \left(A_+ + \zeta e^{-2|\alpha|t} \mathcal{A}^{(1)}_{+}\right)\rho_+(z_1) + \left(B_+ + \zeta \mathcal B^{(1)}_{+}\right)\rho_-(z_1), \\
   \nonumber &\rho_{-}^M(z_2)\rightarrow \rho_-(z_2) + A_- \rho_+(z_2) + B_- \rho_-(z_2), \\
    &\rho_{-}^M(z_1)\rightarrow \rho_-(z_1) + \left(A_- + \zeta  \mathcal{A}^{(1)}_{-}\right)\rho_+(z_1) + \left(B_- + \zeta e^{2|\alpha|t}\mathcal B^{(1)}_{-}\right)\rho_-(z_1)\,, \\
 \nonumber   & \rho_{+}^M(z_2)\rightarrow \rho_+(z_2) + A_+ \rho_+(z_2) + B_+ \rho_-(z_2). 
\end{align}
Similarly the $\cO_L \cO_M\cO_L\cO_M$-ordering is obtained from the following substitutions
\begin{align}
  \nonumber  &\rho^M_{+}(z_1)\rightarrow \rho_+(z_1) + A_+ \rho_+(z_1) + B_+ \rho_-(z_1), \\
  \nonumber  &\rho^M_{-}(z_2)\rightarrow \rho_-(z_2) + \left(A_- + \zeta  \mathcal{A}^{(2)}_{-}\right) \rho_+(z_2) + \left(B_- + \zeta e^{2|\alpha|t} \mathcal{B}^{(2)}_{-}\right) \rho_-(z_2), \\
    &\rho^M_{-}(z_1)\rightarrow \rho_-(z_1) + A_- \rho_+(z_1) + B_- \rho_-(z_1)\,, \\
 \nonumber   & \rho^M_{+}(z_2)\rightarrow \rho_+(z_2) + \left(A_+ + \zeta e^{-2|\alpha|t} \mathcal{A}^{(2)}_{+}\right) \rho_+(z_2) + \left(B_+ + \zeta  \mathcal{B}^{(2)}_{+}\right) \rho_-(z_2). 
\end{align}
Earlier we  defined the normalized negative sum \eqref{alternatingterms} of the two alternating terms which we invoke  here for convenience,
\begin{equation}
    \langle \mathcal H|\mathcal H\rangle \, \mathcal F^{(2n,4)}_{H,h}(t) = -\langle \mathcal H|\cO_M \cO_L\cO_M\cO_L|\mathcal H\rangle - \langle \mathcal H|\cO_L \cO_M\cO_L\cO_M|\mathcal H\rangle.
\end{equation}
Expanding the right-hand side in $\zeta\ll1$ and keeping only the terms that grow at late times $t\gg 1$, the light operator sector becomes,
\begin{equation}\label{F-rho}
    \mathcal F^{(2n,4)}_{H,h}(t) \propto \left(\rho_{+}(z_1)\rho_-(z_2)-\rho_{+}(z_2)\rho_-(z_1)\right)^{-2h_L}\left( 1 +2 h_L \zeta e^{2|\alpha| t}\mathcal{B}^{(1)}_{-}(t)\right)\,.
\end{equation}
Equating \eqref{F-rho}  to \eqref{Lyapunov} we can read off that the Lyapunov exponent given by
\begin{tcolorbox}[colback=yellow!7, colframe=blue!60!black]
\begin{equation}\label{L-a}
    \lambda = 2|\alpha|.
\end{equation}
\end{tcolorbox}
Since in \cite{Vos:2025voq} it was concluded that the temperature is related to the Floquet exponent through $\beta = \frac{\pi}{|\alpha|}$ we can eliminate $\alpha$ and thus \eqref{L-a} yields
\begin{equation}
    \lambda = \frac{2\pi}{\beta}.
\end{equation}
We conclude that the Lyapunov exponent exactly saturates the MSS-bound. We can also read off the scrambling time $t_*$ by observing that since $B^{(1)}_{-}(t)$ is approximately of order unity, the timescale $t_*$ is such that $2h_L\zeta e^{2|\alpha|t} \approx \mathcal{O}(1)$. Reinstating $\zeta = h_M/c$ and eliminating $\alpha$ in favor of the temperature we find that
  \begin{equation}\label{s-t}
    t_* \approx \frac{\beta}{2\pi} \log\left(\frac{c}{2h_Lh_M}\right).
\end{equation}  
This conclusion leads to the intuition that was discussed in the introduction section. In \cite{Banerjee:2018tut,Vos:2025voq} it was found that the state $|\mathcal H\rangle$, while a time-dependent state, introduces a long-lived mode that drives the probe two-point function to its thermal expectation value. We now conclude that the same mechanism drives the four probe OTOC to its chaotic values. The resulting Lyapunov exponent saturates the MSS-bound with the effective temperature set by the Floquet exponent. This is what we would expect for holographic CFTs dual to semi-classical gravity. However we have shown that the bound is saturated only to the leading $c$- correction. As it was emphasized in \cite{Fitzpatrick:2016thx,Chen:2016cms} we would expect subleading corrections in \eqref{L-a} which were not considered in our setting.

\section{Discussion and outlook}

\label{sec:discussion}

In this work we investigated chaotic dynamics within a class of time-dependent states of two dimensional CFT. To do so, we computed an out of time ordered correlator (OTOC) of four light operators within a time-dependent pure state of the heavy operators by employing the monodromy method. 

A key step in extracting this quantity was to promote two of the four light operators entering the OTOC into middle-heavy operators, interpreted as perturbations on top of the heavy background, yielding a hierarchy of scales in respective operator dimension with the medium-heavy operators taken to have conformal dimension much smaller than those of the heavy operators but much larger than those of the light operators. This splitting introduced the essential coupling within the light sector which permitted the extraction of the exponent, which could not be done from analyzing the leading order light sector contributions alone where all the light operators were treated on the same footing, manifesting as two-point function factorization of the four-point function. This coupling was realized through the presence of branch discontinuities which emerged when circling the light operators around the middle-heavy ones, a process exemplifying the out of time ordered nature of the correlator and solidifying a proper ordering prescription for the operators themselves. To then establish the correct time dependence of the entire system and hence dynamics and scrambling, a correct analytic continuation to Lorentzian signature was performed. Exploiting Floquet evolution of the particular function basis entering the expressions of the correlators, an association between its characterizing exponential parameter and system temperature was made, subsequently allowing for the Lyapunov exponent to be identified according to the prescribed exponential decay of the correlator, explicitly realizing the MSS bound in our time-dependent setup. 

Note in \eqref{ourOTOC}, the presence of the bounded function $\mathcal B^{(1)}_-(t)$. It does not affect the extraction of the Lyapunov exponent since the scrambling time is significantly longer than the period of $\mathcal B^{(1)}_-(t)$. It mirrors the coarse-graining discussion in \cite{Vos:2025voq}. In our case the leading exponential behavior was multiplied by a periodic function and hence the leading exponential behavior, which only became visible at timescales much longer than the period of periodic function.

 In this work, notably, we focused entirely on the dynamics within two-dimensional CFT. However, given the well studied AdS$_{3}$/CFT$_2$ holographic dictionary, it is natural to ask to what do these dynamics correspond in the gravitational dual. It would be interesting to interpret our results as potentially capturing late time dynamics of gravitational collapse from initial configurations of multi-centered black holes which has been investigated in various aspects in \cite{Hulik:2016ifr,Hulik:2019pwr}.

Furthermore, here and in \cite{Vos:2025voq} the focus lay on the leading exponential decay of the OTOC and the two-point expectation value respectively. In the case where the heavy state consists of a high-energy pure state it can be verified that the identity block yields the full thermal two-point function \cite{Fitzpatrick:2015zha}. The thermal two-point function not only predicts the temperature but also contains knowledge of the higher quasi-normal mode spectrum \cite{Birmingham:2001pj,Horowitz:1999jd}. Given the earlier speculative connection to bulk gravitational collapse it would be interesting to see whether some information could be extracted about the ringdown spectrum.

Finally, there are many tangential directions that can be pursued as well. Instead of working  only with the holomorphic sector, one can also include a non-trivial anti-holomorphic sector which corresponds   to states with angular momentum. As  \cite{Jahnke:2019gxr} suggests  there should be two separate Lyapunov exponents for the  effective temperatures of the left- and right-moving sectors. Furthermore, since in our setup the initial and final state are related by conjugation there is a manifest reflection symmetry in the heavy sector. This symmetry was used to prove that the Hill potential was real-valued and it is  similar to the doubling trick utilized in boundary CFT. It would be interesting to see if our methods carry over to boundary CFT as well.


\section*{Acknowledgements}

We would like to thank Shiraz Minwalla, Sungjay Lee and Ren\'e Meyer for useful discussion. We also thank Kyriakos Papadodimas for inspiring this project. The work of SDH is supported by KIAS Grant PG096301. The work of PP is supported by KIAS Grant PG105601. The work of GV is supported by KIAS Grant PG PG100401. GV would also like to thank CEICO and the University of W\"urzburg for their hospitality the period in which this work was completed. We would also like to thank the participants of the ELB workshop at KIAS and the 9th Holography, String Theory and Observational Cosmology workshop at Chiang Mai for useful feedback.

\appendix

\section{Converting the branch discontinuity to real-time parametrization}

\label{sec:realtimeappendix}

In this appendix we provide the exact expressions for the accessory parameters and branch discontinuities in terms of the real-time Floquet parametrization. The results of these calculations are quoted in section \ref{sec:quantumchaos}. On the original Euclidean radial plane the branch discontinuity terms are given by 
\begin{align}
  \nonumber  & \frac{1}{2\pi i} \oint_{\mathcal C_1} dz A'_{\pm}= -\frac{6}{c} c_1\, \rho_{\pm}(w_1)\rho_+(w_1) - 6 \zeta \left(\rho_{\pm}'(w_1)\rho_+(w_1) + \rho_{\pm}(w_1)\rho_+'(w_1)\right),  \\
    & \frac{1}{2\pi i} \oint_{\mathcal C_1}dz B'_{\pm}= \frac{6}{c} c_1\, \rho_{\pm}(w_1)\rho_-(w_1) + 6\zeta \left(\rho_{\pm}'(w_1)\rho_-(w_1) + \rho_{\pm}(w_1)\rho_-'(w_1)\right)\,, 
\end{align}
and 
\begin{align}
  \nonumber  & \frac{1}{2\pi i} \oint_{\mathcal C_2}dz A'_{\pm}= \frac{6}{c} c_2 \,\rho_{\pm}(w_2)\rho_+(w_2) + 6 \zeta \left(\rho_{\pm}'(w_2)\rho_+(w_2) + \rho_{\pm}(w_2)\rho_+'(w_2)\right), \\
    & \frac{1}{2\pi i} \oint_{\mathcal C_2} dz B'_{\pm}= -\frac{6}{c} c_2\, \rho_{\pm}(w_2)\rho_-(w_2) - 6\zeta \left(\rho_{\pm}'(w_2)\rho_-(w_2) + \rho_{\pm}(w_2)\rho_-'(w_2)\right)\,. 
\end{align}
where  the accessory parameters $c_1$ and $c_2$ are given by
\begin{align}
    & c_1 = -2h_M  \frac{\rho_+'(w_1)\rho_-(w_2)-\rho_+(w_2)\rho_-'(w_1)}{\rho_+(w_1)\rho_-(w_2)-\rho_+(w_2)\rho_-(w_1)}, \label{appaccesoryparameter1}\\
    & c_2 = -2h_M  \frac{\rho_+(w_1)\rho_-'(w_2)-\rho_-(w_1)\rho_+'(w_2)}{\rho_+(w_1)\rho_-(w_2)-\rho_+(w_2)\rho_-(w_1)}. \label{appaccesoryparameter2}
\end{align}
As was established in section \ref{sec:quantumchaos}, the parametrized trajectory on the complex plane that corresponds to turning on Lorentzian time for the middle heavy operators corresponds to
\begin{equation}
    w_1(t) = e^{2i(t+i\eps)}, \qquad w_2(t) = e^{2i(t-i\eps)}. 
\end{equation}
The goal is to re-express the above formulas in a form that shows the dependence on $t$ in a more explicit manner. To accomplish this, as explained in section \ref{sec:quantumchaos}, we parametrize the argument of $\rho_{\pm}(w)$ by $w = e^{2is}$ with $s\in \mathbb{R}$ and rewrite $\rho_{\pm}(w)$ in terms of the Floquet basis,
\begin{equation}
    \rho_{\pm}(e^{2is}) = \sqrt{2i}  e^{(\mp |\alpha| +i)s}p_{\pm}(s),
\end{equation}
where $p_{\pm}(s)$ is a periodic function of $s$, namely $p(s+\pi)=p(s)$. We will assume that  we can analytically continue the real domain of $p_{\pm}(s)$ to include a thin strip surrounding the real axis so that we can evaluate $p_{\pm}(s)$ at $s=t\pm i\eps$. In order to express the branch discontinuities we need the derivatives and the accessory parameters. The derivatives are easily expressed through means of the chain rule resulting in
\begin{equation}
    p_{\pm}'(s) = \frac{1}{\sqrt{2i}}\left((\mp|\alpha| +i)e^{(\mp|\alpha|-i)s}p_{\pm}(s) + e^{(\mp|\alpha|-i)s)}p_{\pm}'(s)\right).
\end{equation}
Note that the derivatives also factorize into a product of an exponential factor and a bounded factor and we  define,
\begin{equation}
    p_{\pm}'(s) = \frac{1}{\sqrt{2\pi i}} e^{\mp|\alpha|s}q_{\pm}(s).
\end{equation}
These derivatives can be substituted into the accessory parameters. In order to condense the notation we also define the following 
\begin{equation}
    t_{\eps} \equiv t + i\eps.
\end{equation}
In this case, the accessory parameters, as functions of time,  read
\begin{align}
\nonumber c_1 = &\quad\frac{e^{-2it_{\eps}}\, H_m \Big(\, i\, e^{4 i |\alpha| \epsilon}\, p_+(t_{-\eps}) \big( (i + |\alpha|)\, p_-(t_{\eps}) + p_-'(t_{\eps}) \big)\Big)}{e^{4 i |\alpha| \epsilon}\, p_-(t_{\eps})\, p_+(t_{-\eps}) - p_-(t_{-\eps})\, p_+(t_{\eps})}
      \\[9pt]
      &+\frac{e^{-2it_{\eps}}\, H_m\Big( p_-(t_{-\eps}) \big( (1 + i |\alpha|)\, p_+(t_{\eps}) - i\, p_+'(t_{\eps}) \big) \Big)}
     {e^{4 i |\alpha| \epsilon}\, p_-(t_{\eps})\, p_+(t_{-\eps}) - p_-(t_{-\eps})\, p_+(t_{\eps})}\,,
\end{align}
and similarly 
\begin{align}
\nonumber c_2 = &\quad\frac{e^{-2it_{-\eps}}\, H_m \Big(\, p_+(t_{\eps}) \big( (1 - i |\alpha|)\, p_-(t_{-\eps}) - i\, p_-'(t_{-\eps}) \big)\Big)}
     {e^{4 i |\alpha| \epsilon}\, p_-(t_{\eps})\, p_+(t_{-\eps}) - p_-(t_{-\eps})\, p_+(t_{\eps})}\\[9pt]
    &-\frac{e^{-2it_{-\eps}}\, H_m \Big( i\, e^{4 i |\alpha| \epsilon}\, p_-(t_{\eps}) \big( (|\alpha| - i)\, p_+(t_{-\eps}) + p_+'(t_{-\eps}) \big) \Big)} {e^{4 i |\alpha| \epsilon}\, p_-(t_{\eps})\, p_+(t_{\eps}) - p_-(t_{-\eps})\, p_+(t_{\eps})}\,.
\end{align}
The significant conclusion is that the exponentially growing and decaying factors cancel each other in such a way that the accessory parameters become bounded functions of time. The branch discontinuities can now be expressed as well. Since there are two relevant contours, two coefficients $A, B$ in the variation of parameters expansion and two subscripts $\pm$, there are eight individual branch discontinuities in total. Before we express them explicitly we can identify a pattern in the overall exponential factors. They can either cancel or enhance each other so it can be seen that they can be classified as in \eqref{disc-behavior}. Explicitly, the branch discontinuities are succinctly written as follows,
\begin{align}
\nonumber\frac{1}{2\pi i} \oint_{\mathcal C_1} dz A'_+ &= -12\, e^{-2|\alpha| t_{\epsilon}}\, p_+(t_{\epsilon}) \Big[ \Big( i\frac{c_2}{c}\, e^{2 i t_{\epsilon}} + (i - |\alpha|)\zeta \Big) p_+(t_{\epsilon}) + \zeta\, p_+'(t_{\epsilon}) \Big]\,, \\
\nonumber\frac{1}{2\pi i} \oint_{\mathcal C_1}dz A'_- &= -6\zeta\, p_+(t_{\epsilon})\, p_-'(t_{\epsilon}) - 6\, p_-(t_{\epsilon}) \Big[ 2i\Big( \frac{c_2}{c}\, e^{2 i t_{\epsilon}} + \zeta \Big) p_+(t_{\epsilon}) + \zeta\, p_+'(t_{\epsilon}) \Big]\,, \\
\nonumber\frac{1}{2\pi i} \oint_{\mathcal C_1}dz B'_+ &= 6\zeta\, p_+(t_{\epsilon})\, p_-'(t_{\epsilon}) + 6\, p_-(t_{\epsilon}) \Big[ 2i\Big( \frac{c_2}{c}\, e^{2 i t_{\epsilon}} + \zeta \Big) p_+(t_{\epsilon}) + \zeta\, p_+'(t_{\epsilon}) \Big]\,, \\
\frac{1}{2\pi i} \oint_{\mathcal C_1} dz B'_- &= 12\, e^{2|\alpha| t_{\epsilon}}\, p_-(t_{\epsilon}) \Big[ \Big( i\frac{c_2}{c}\, e^{2 i t_{\epsilon}} + (i + |\alpha|)\zeta \Big) p_-(t_{\epsilon}) + \zeta\, p_-'(t_{\epsilon}) \Big]\,,
\end{align}
and 
\begin{align}
\nonumber\frac{1}{2\pi i} \oint_{\mathcal C_2} dz A'_+ &= 12\, e^{-2|\alpha| t_{-\epsilon}}\, p_+(t_{-\epsilon}) \Big[ \Big( i\frac{c_2}{c}\, e^{2 i t_{-\epsilon}} + (i - |\alpha|)\zeta \Big) p_+(t_{-\epsilon}) + \zeta\, p_+'(t_{-\epsilon}) \Big]\,, \\
\nonumber\frac{1}{2\pi i} \oint_{\mathcal C_2} dz A'_- &= 6\zeta\, p_+(t_{-\epsilon})\, p_-'(t_{-\epsilon}) + 6\, p_-(t_{-\epsilon}) \Big[ 2i\Big(\frac{c_2}{c}\, e^{2 i t_{-\epsilon}} + \zeta \Big) p_+(t_{-\epsilon}) + \zeta\, p_+'(t_{-\epsilon}) \Big]\,, \\
\nonumber\frac{1}{2\pi i} \oint_{\mathcal C_2}dz B'_+ &= -6\zeta\, p_+(t_{-\epsilon})\, p_-'(t_{-\epsilon}) - 6\, p_-(t_{-\epsilon}) \Big[ 2i\Big( \frac{c_2}{c}\, e^{2 i t_{-\epsilon}} + \zeta \Big) p_+(t_{-\epsilon}) + \zeta\, p_+'(t_{-\epsilon}) \Big], \\
\frac{1}{2\pi i} \oint_{\mathcal C_2}dz B'_- &= -12\, e^{2|\alpha| t_{-\epsilon}}\, p_-(t_{-\epsilon}) \Big[ \Big( i\frac{c_2}{c}\, e^{2 i t_{-\epsilon}} + (i + |\alpha|)\zeta \Big) p_-(t_{-\epsilon}) + \zeta\, p_-'(t_{-\epsilon}) \Big]\,.
\end{align} 
As stated, the branch discontinuities still factorize into the product of an exponential part and a bounded part.  As stated in the main text the discontinuities  can be cast in the form \eqref{AB-behavior}.
Quite importantly, by construction, the middle-heavy accessory parameters were derived in such a way that there is no branch discontinuity along a loop that goes around both  $w_1$ and $w_2$ points. This implies that  the branch discontinuities cancel each other, namely 
\begin{align}
    & \oint_{\mathcal C_1}dz  A'_{\pm} - \oint_{\mathcal C_2} dz A'_{\pm} =0, \qquad \oint_{\mathcal C_1} dz B'_{\pm} - \oint_{\mathcal C_2}dz B'_{\pm}= 0,
\end{align}
which is  due to the relative orientation change between $\mathcal C_1$ and $\mathcal C_2$. From this we obtain the identities \eqref{AB-constraints},
\begin{align}
    & \mathcal{A}^{(1)}_{\pm} = \mathcal{A}^{(2)}_{\pm}, \qquad \mathcal{B}^{(1)}_{\pm} = \mathcal{B}^{(2)}_{\pm}.
\end{align}
which can be verified from the Wronskian normalization condition \eqref{Wronskian}.

\renewcommand{\thesection}{\Alph{section}}

\renewcommand{\theequation}{\Alph{section}.\arabic{equation}}

\setcounter{section}{0}


\bibliographystyle{JHEP}
\bibliography{refs}

\end{document}